# Dynamics of femtosecond synthesized coronary profile laser beams filamentation in air

Yu.E. Geints*, A.A. Zemlyanov

V.E. Zuev Institute of Atmospheric Optics SB RAS, Zuev square 1, 634021, Tomsk, Russia
*Corresponding author: ygeints@iao.ru

## Abstract

Multiple filamentation in air of high-power ultrashort laser radiation with transverse intensity profile resembling a "corona" composed by incoherent combining of several annularly distributed independent top-hat sub-beams is theoretically studied. Through the numerical solution of time-averaged nonlinear Schrodinger equation, we study the spatio-angular dynamics of synthesized near-infrared "corona-beam" (CB) along the optical path by varying the number and power of the beamlets (corona-spikes). For the first time to our knowledge, the evident advances in the multiple-filamentation region manipulating of synthesized CB are demonstrated. Particularly, by adjusting the number and aperture of the constituting sub-beams it makes possible to significantly delay the CB filamentation onset distance and increase the filamentation length in air. In addition, at the post-filamentation stage of femtosecond pulse propagation under certain conditions the synthesized corona-beams exhibit significantly lower angular divergence of its most intense part (post-filamentation light channel) compared to the beams with regular unimodal intensity profiles (Gaussian, plateau-like) that provides enhancing of laser power delivered to the receiver over the atmospheric links.

## 1. Introduction

The phenomenon of light self-focusing in various optical media has been actively investigated since the 1960s [1-5]. Close attention to this problem is connected with the fact that the laser pulse self-focusing is a bright manifestation of the nonlinear physics effects and possesses important practical potential for high-power atmospheric optics [6-8]. As the peak power of an optical radiation (typically an ultrashort laser pulse) exceeds some critical value, on the propagation path a self-induced convex "lens" is formed due to the action of the optical Kerr effect which focuses the radiation. As a result of this nonlinear and usually aberrational self-focusing, a narrow high-intensity light filament (or group of filaments) is formed within the pulse. This process is accompanied by significant enrichment of the pulse spectral composition as a result of strong self-phase modulations in a nonlinear medium. This is manifested in the formation of wide supercontinual wings covering in some cases several octaves of the initial pulse spectrum.

High intensity of the optical field inside the filaments causes multiphoton and tunnel ionization in the medium, which leads to plasma generation in the beam wake with a characteristic density of free electrons in the range from $10^{14}$ to $10^{18}$ cm$^{-3}$ [1]. The visual indicator of laser pulse filamentation is the luminescence of these plasma channels in the visible spectrum due to the recombination of plasma electrons with neutral molecules. Usually, the distance of laser pulse filamentation is unambiguously associated with the region of this plasma channels existence.

One of the main challenges in practical applications of wide-aperture laser radiation filamentation on atmospheric links is the control over the spatial position and spatial structure of the filamentation region. To this end, the use of specially profiled radiation, i.e. laser beams with non-Gaussian transverse intensity distribution, seems to be very promising. So far, there are several studies reported in the literature, when various structured laser beams filamentation are considered, e.g., the dark-hollow (ring) beam [9], super-Gaussian profile [10], (quasi) diffraction-free Bessel-Gaussian beams [11, 12], as well as the so-called "dressed" beam [13, 14] obtained by superposition of a Gaussian and ring-beam profiles. Practical interest in profiled radiation is related to the specific features of linear diffraction of such beams, which in turn opens up prospects for additional control over the nonlinear propagation region, i.e., the filamentation area.

The obtaining of complicated intensity distribution is an independent and challenging technical task requiring controlled distortions of amplitude and phase of initial radiation using various amplitude masks, phase transparencies, segmented and flexible mirrors [11, 15-19] etc.. It should be noted that in all the above-mentioned studies, the technique of introducing special pre-distortions into the initial and as a rule unimodal laser beam profile is used. Then, this distorted beam during the propagation within the Fresnel zone can form the spatially coherent regions with predominant localization of pulse power. Exactly these areas serve as the sources of laser filamentation under the influence of beam self-focusing.

At the same time, a certain alternative here is the technique proposed by the scientific group from the U.S. Navy Laboratory [20, 21] of incoherent combining the low-power and narrow aperture fiber lasers into a single wide-aperture radiating matrix, giving the resulting laser beam of the higher power required. Such a synthesized beam in the form of a matrix of four emitters is experimentally proved to significantly increase the potential of long-range atmospheric systems for laser energy transmission (> 3 km) without the use of any additional adaptive optics in the conditions of environmental jamming introduced by atmospheric turbulence and aerosol scattering [21]. It should be emphasized that in these experiments the incoherent sub-beams combining is used without any need of a complicated and expensive control over the partial phases of sub-apertures [22].

To date, in the extensive literature on the physics of femtosecond laser filamentation, the examples of the use of this type of synthesized laser radiation cannot be found. As a rule, complex multimodal intensity distributions are obtained from a single laser beam by modulating its amplitude-phase profile [11, 18], or by the optical partitioning into several spatially separated coherent sub-beams with decreased power [23, 24]. In this connection, it is important studying the laser pulse filamentation dynamics of synthesized beams with incoherently emitting sub-apertures. As shown in this paper, using a specific annular type of the spatial arrangement of incoherent sub-beams makes it possible to gain additional control on the distance and length of the filamentation region, as well as to significantly reduce the angular beam spreading during the post-filamentation propagation in air, thus increasing the optical power delivered through the atmospheric links. Worthwhile noting, the closest related beam spatial profile to the corona-beam studied here is that considered in Ref. [25]. However, this type of field distribution is formed as a result of Kerr-driven collapse of complicated hybrid-polarized vector field with initial ring-type amplitude distribution that is difficult to realize in practice with high-power laser radiation.

## 2. Linear propagation dynamics of synthesized corona-beam.

In the numerical calculations, we consider a certain type of synthesized optical field distribution composed of a bundle of $N_b$ isolated sub-beams with equal amplitude and aperture radius $r_0$ arranged in a circular ring with radius $R_0$ (Fig. 1(a)). Visually, this type of spatial intensity distribution resembles a corona and hereafter we will call it a "corona-beam" (CB). Obviously, in the limit $N_b \gg 1$ a CB transforms into a normal annular dark-hollow beam. The following expression defines the CB initial transverse profile of the electric field $U_{cb}(x, y)$:

$$U_{cb}(x,y) = U_0 \sqrt{\sum_{m=1}^{N_b} |U_m(x_m, y_m)|^2}, \quad x_m = R_0 \cos\left(\frac{2\pi m}{N_b}\right), \quad y_m = R_0 \sin\left(\frac{2\pi m}{N_b}\right), \quad m = 1 \ldots N_p, \quad (1)$$

where $U_0$ is the scaling multiplier, and each partial maximum (lobe) $U_m$ is defined by a super-Gaussian profile with a half-width $r_0$ imitating the radiation of a powerful fiber laser with a spatially limited aperture:

$$U_m(x_m, y_m) = \exp\left\{-\left[(x - x_m)^2 + (y - y_m)^2\right]^s \Big/ 2r_0^{2s}\right\} \quad (2)$$

with $s = 3$. As seen, Eq.(1) defines incoherently combined corona-beam without any phase control between beamlets.

Additionally, we also introduce a modification of CB profile obtained from the superposition of the annular distribution Eq. (1) and a single super-Gaussian beam Eq. (2) placed on the optical axis (Fig. 1(d)):

$$U_{db}(x,y) = \sqrt{U_{cb}^2(x,y) + U_d^2 |U_m(x_m = 0, y_m = 0)|^2}, \quad (3)$$

where the parameter $U_d$ sets the amplitude of the central lobe of synthesized beam. The surface generating function Eq. (3) defines a certain analogue of the above mentioned "dressed" beam distribution [13, 14], in which the outer ring is shattered into isolated sub-beams. Therefore for convenience, the distribution defined by Eq. (3) is referred here to as a "dressed" coronary beam (DCB).

It is instructive to trace how the transverse profile of the considered beam types is transformed during free propagation in air. The results of such linear diffraction simulation without accounting for any nonlinear and chromatic dispersion effects in the propagation medium are shown in Figs. 1(a)-(i). The numerical calculations are carried out based on the parabolic equation for the complex electric field amplitude $U(x,y)$ of optical radiation (see, Eq. (A1) in Appendix). Here and hereafter, the initial radius of the synthesized beam is set to $R_0 = 5$ mm and the optical carrier wavelength is $\lambda_0 = 800$ nm. In the figures, the aperture radius of the partial sub-beams is $r_0 = 0.3R_0$ and their number $N_b$ equals to 8.

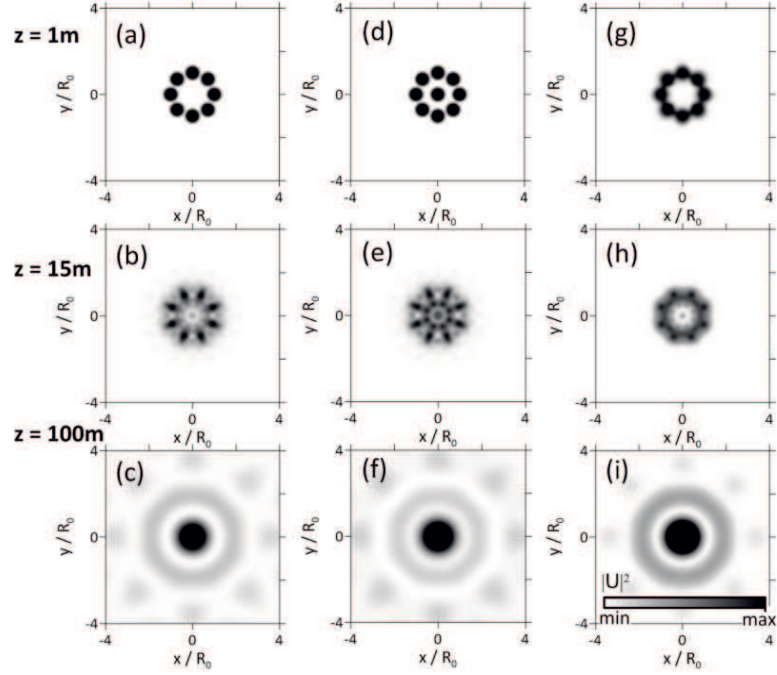

Fig. 1. Transverse intensity profiles $|U|^2$ of CB (a-c) and DCB (d-f) with $N_b = 8$ during linear propagation in air to different distances; (g-i) coherently combined octagon corona-beam.

From Figs. 1(b), 1(d) is clear that already after approximately 15 m from the optical path beginning, the synthesized corona-beams evolve to spatially complicate light structures consisting of many lobes coupled with intensity rings as a result of field interference of individual sub-beams. The DCB diffraction patterns differ from the CB profiles by a strongly pronounced central maximum and a better developed annular intensity structure. Note that the considered distance ($z = 15$ m) corresponds to the near-field diffraction region, because it is close to the Rayleigh length of the considered beams, $L_R = \pi R_0^2 / (\lambda_0 M^2) \approx 16$ m, where the beam quality parameter [26] $M^2 \approx 6$ for both combined distributions (CB and DCB).

In the far-field region $z \gg L_R$ (Figs. 1(c, e)) the differences in the diffraction patterns of the two types of synthesized beams practically disappear and their profiles exhibit the leading intensity maximum located at the intensity distribution centroid (on the optical axis), which is surrounded by a weaker intensity ring and $N_b$ secondary lobes. The calculations show that a similar field structure of the synthesized beams in the far-field region is typical for any sub-beams annular compositions.

For comparison, Figs. 1(g)-(i) show the linear diffraction in air of the coronary profiled beam with coherently combined sub-beams. Since the partial emitting sub-apertures are located close to each other, their mutual field interference becomes noticeable already close to the propagation onset ($z = 1$ m). This leads to strongly pronounced blurring of beam transverse diffraction profile in the near diffraction region ($z < L_R$) without distinct separation of intensity maxima corresponding to individual emitters that is typical for incoherently combined beams bundle (see, Figs. 1(b), (e)). Although, in the far-field (Fig. 1(i)) such a coherently synthesized beam becomes structurally similar to its counterpart with incoherent sub-beams mixing, the filamentation dynamics of these beams is quite different because the self-focusing starts in the Fresnel zone where the beam profiles differ substantially.

### 3. Filamentation dynamics of high-power synthesized corona-beams in air

Now consider the nonlinear propagation dynamics of synthesized multimodal laser pulses with peak power sufficient to excite nonlinear-optical effects of self-focusing and filamentation in air. As known, because of filamentation, such high-power laser radiation experiences strong spatial and temporal self-phase modulation that results in large-scale changes in the spectral pulse composition, beam spatial partitioning into high-intensity areas, and the formation of high-density plasma channels along the pulse propagation path [1-3].

Commonly, for the theoretical description of filamentation dynamics of high-intense ultra-short laser radiation the (3+1)-dimensional nonlinear Schrodinger equation (NLSE) is used [1]. As an alternative to NLSE, the more general spectral analogue in the form of the unidirectional pulse propagation equation (UPPE) is known [27], which is capable for resolving the carrier of the complex optical field in the spatio-temporal frequencies domain. Due to the extremely large volume of computational operations required for the numerical solution of a full four-dimensional model in the framework of NLSE or UPPE, we used the reduced version of NLSE formulated in the 3D-space to simulate the self-action of a wide-aperture (centimeter) laser beam. This approach relies on the time-integrated NLSE and was first proposed in [28] for the numerical simulation of Ti:Sa-laser beam multiple filamentation in air where it demonstrated good quantitative agreement on the nonlinear focus position and length of filamentation with respect to the full (3+1)-dimensional problem taking into account the temporal structure of the pulse. The specific form of this reduced NLSE is considered in detail in [29] and an example of numerical simulations of CB filamentation in comparison with the full model are given in Appendix.

The results of the nonlinear propagation simulation of three CBs with equal initial peak power $P_0$ being ten times higher than the critical power of self-focusing $P_0$ (the parameter of reduced power $\eta \equiv P_0/P_c = 10$), and different number $N_b$ of sub-beams are shown in Figs. 2(a-l) in the notation "$CB_{N_b}$". Here, the transverse distributions of relative optical intensity $I/I_0$ are plotted, where $I = |U|^2$, at some selected distances along the propagation path. Note that due to the condition of constant peak power $P_0$ of CB with different number $N_b$ of sub-beams within the synthesized aperture, their initial size $r_0$ and intensity $I_0$ were also different: $r_0 = 0.3 R_0$, $I_0 = 0.12$ TW/cm$^2$ (CB$_4$), $r_0 = 0.3 R_0$, $I_0 = 0.06$ TW/cm$^2$ (CB$_8$) and $r_0 = 0.2 R_0$, $I_0 = 0.1$ TW/cm$^2$ (CB$_{12}$).

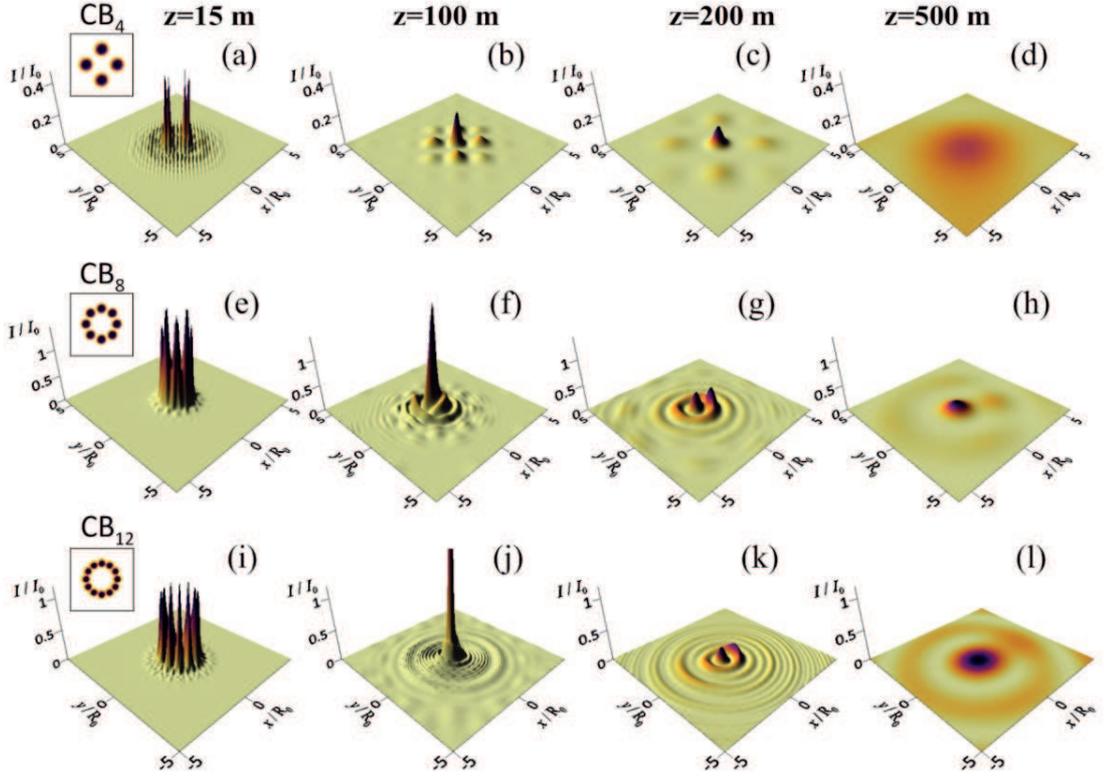

Fig. 2. Normalized intensity distribution of corona-beams with different sub-beams number $N_b = 4$ (a)-(d), 8 (e)-(h), 12 (i)-(l) and reduced power $\eta = 10$ at different distances in air as indicated by row header: $z = 15$ m, 100 m, 200 m, 500 m.

The trace evolution of relative intensity and maximal value of free electrons density $\rho_e$ in the photoionization plasma are presented in Figs. 3(a, b) upon nonlinear propagation in air of different corona-beams as well as a super-Gaussian beam (denoted as "SG" in the figures) with the profile defined by Eq. (2), but with initial aperture radius $r_0 = R_0$, i.e., with the full radius of the synthesized corona-beam.

As seen, all the considered beams demonstrate self-focusing at some distance from the beginning of the propagation resulting in a sharp increase of pulse intensity and active plasma generation in medium reaching the value of $\rho_e \approx 10^{16}$ cm$^{-3}$. Meanwhile, from Fig. 3(c) it follows that the self-focusing distance $z_{sf}$ increases with increasing number of corona sub-beams, from $z_{sf} \approx 8$ m for CB$_4$ to about 80 m for CB$_{12}$ and $\eta = 10$.

Importantly, for CB with $N_b = 12$ and ten-fold excess of the critical power ($\eta = 10$), the filamentation is also observed despite the apparent power shortage in each sub-beam: $\eta_b = \eta/N_b < 1$. In other words, for the incoherently combined beam as a whole, the filamentation is realized even when the partial peak power of each sub-beam becomes below the critical value for its self-focusing ($\eta_b < 1$), and the transverse collapse of each sub-beam due to the Kerr nonlinearity should be stopped by strong sub-beam diffraction.

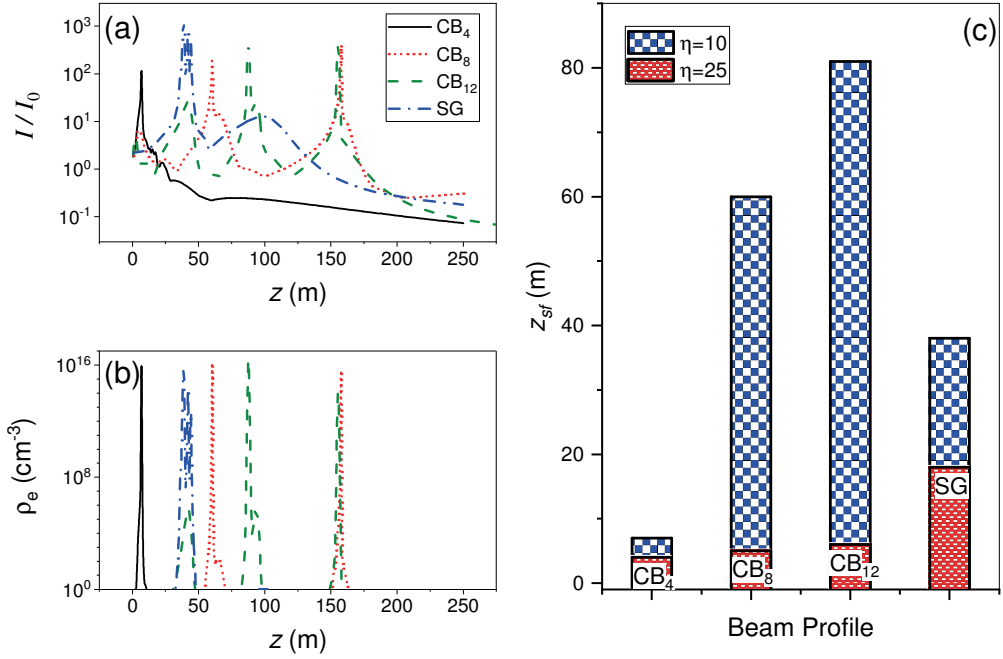

Fig. 3. (a) Relative intensity $I/I_0$, (b) peak plasma electrons density $\rho_e$ and (c) self-focusing distance $z_{sf}$ for laser beams with different spatial profile and initial power.

Interestingly, a similar scenario with two subcritical $\eta_b < 1$ beams separated in space was reported earlier in [23, 30] when this twin beam may undergo filamentation as a result of sub-beams diffractive "fusion", which leads to net power increase of the common field structure above the critical level ($2\eta_b > 1$). It is this situation that is realized in the case of $CB_{12}$ with the only difference that additionally to the condition for a certain minimum distance between sub-beams, their mutual configuration in space is also of importance.

Indeed, recalling the beam profiles in Figs. 1 and 2, we note that the circular arrangement of the partial sub-beams in corona-shaped aperture always leads to the formation of the diffraction maximum in the centroid of intensity distribution with any $N_b$ number. This central maximum becomes leading already at the distances close to beam Rayleigh length $L_R$ and further experiences filamentation. At the same time, if the number of sub-beams in the synthesized beam is small but their initial power is high enough ($\eta_b \gg 1$) as in the case of $CB_4$ (Figs. 2(a-d)), the filamentation may also start separately in each sub-aperture. This explains a significant decrease in the self-focusing distance for such a beam profile (see, Fig. 3(c)). Worthwhile noting, the filamentation in the axial maximum rather than of the sub-beams provides for high $z_{sf}$ values for all CBs considered. Moreover, this distance is even greater than that of a unimodal beam of super-Gaussian profile (SG) with the initial radius $R_0$, when SG self-focusing distance reaches $z_{sf} \approx 40$ m.

Additional peculiarity of corona-beams filamentation should be noted. As shown in Figs. 2(f) and 2(j), after exiting the primary nonlinear focus located at distance $z \cong 100$ m, CB transverse profile consists of one or several axial intensity maxima surrounded by annual ripples. A similar beam profile is observed in the filamentation of a dark-hollow as well as of "dressed" optical beams [9, 14]. Meanwhile, as shown in [13], a certain part of laser pulse energy contained in the annular area serves as a source for energy replenishment of axial (central) filaments and can promote the elongation of the filamentation region. Moreover, in addition to playing a purely "energy" function of filament replenishment, the circular beam area makes a certain impact on the filamentation of central beam region due to the very diffraction specificity of optical radiation with the multimodal profile. This diffraction specificity is originated from the presence of isolated wave structures (central maximum, annual rings), which interfere and experiences strong coupling upon beam propagation in medium, thus affecting the entire course of radiation self-action [14].

Thus, from Figs. 3(a, b) it follows that at moderate pulse power, $\eta = 10$, and the number of sub-apertures in the synthesized beam $N_b = 8$ and 12 the filaments are observed up to 150 m from the propagation beginning. This elongation of the filamentation region does not occur uniformly along the propagation path but in disruptive manner as recurrent beam refocusing producing enhanced intensity and density of plasma electrons. Moreover, for $N_b = 4$ when the effect of common ring structure encompassing the corona-spikes does not occur (see, Figs. 2(a-d)), the filaments are evident only at the beginning of the propagation distance, up to $z \approx 8$ m, at the same initial pulse power.

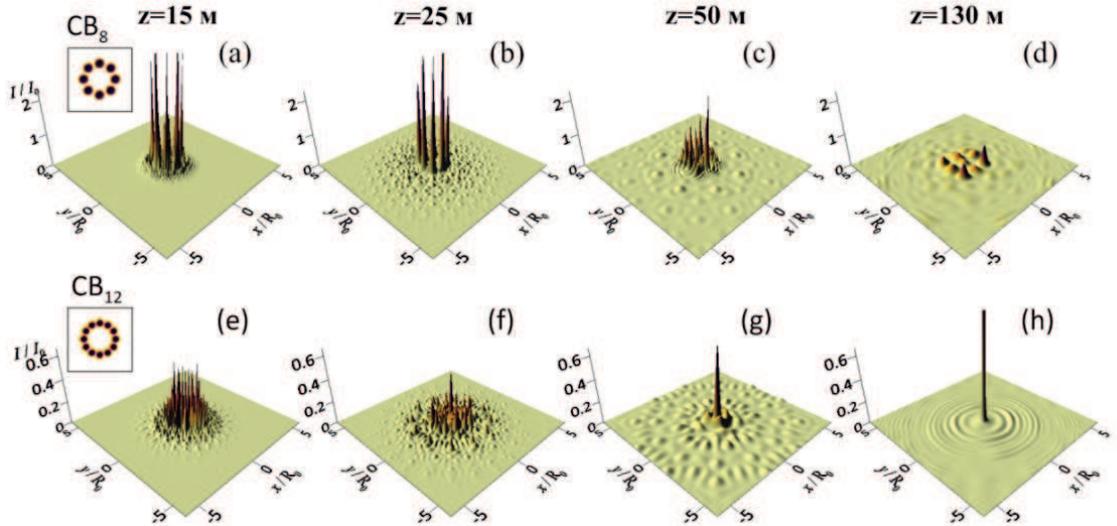

Fig. 4. Transverse intensity distribution of different CBs with reduced power $\eta = 25$ and $N_b = 8$ (a-d) and 12 (e-h) at different ranges in air.

The increase in initial pulse power to $\eta = 25$ affects the dynamics of CB filamentation with different number of sub-apertures in a distinct way. For example, in the combined profile $CB_8$ every corona-spike possesses more than $3 P_c$. This power is sufficient to counteract the diffraction of individual sub-apertures (Figs. 4(a-d)) and their fusion, which occurs with a beam in the form of $CB_{12}$ profile (Figs. 4(e-h)) when every sub-beam contains one and a half times less power. In the case with $CB_8$, the sub-beams self-focus and enter the filamentation conditionally independently, forming a well-defined multimodal profile of nine separate lobes in the far-field region (Fig. 4(d)). In the corona-beam with twelve spikes ($CB_{12}$), the initial sub-beams self-combine into an integral quasi-unimodal distribution with the leading axial maximum surrounded by low-intensity rings to the distance $z = 50$ m (Figs. 4(g, h)).

At the same time, as follows from Fig. 3(c), in contrast to the above considered case with $\eta = 10$, the self-focusing of more powerful corona-beams occurs at the very beginning of the optical path. Here, the self-focusing distance $z_{sf}$ depends on the partial spike power $\eta_b$ and obeys the simple functional relationship obtained in the case of non-aberration Gaussian beam self-focusing (Talanov's formula [31]): $z_{sf} \propto (\eta_b - 1)^{-1/2}$.

### 4. Optical diffraction-rays model of corona-beam filamentation

In this section, the filamentation dynamics of synthesized corona-profile laser beams is considered based on the so-called amplitude-phase approach to laser radiation propagation in nonlinear medium. To this end, we use the concept of diffraction-optical rays and associated diffraction-ray tubes (DRT). The mathematical fundamentals of the diffraction-rays methodology for optical beam propagation is presented in detail, e.g., in [32-34]. This approach helps to visualize a diversity of physical peculiarities accompanying the interaction of high-intensive optical radiation with a nonlinear medium. Many of these features are associated with the transformation of the optical wave phase and cannot be traced within the framework of the regular "pure amplitude" treatment of light pulse evolution.

It should be recalled that the geometric optics (GO) usually visualizes the wavefront based on the technique of geometric rays tracing [35]. The GO-rays are straight lines (in a homogeneous medium) directed along the normal to the wavefront in every spatial point. For optical beam of finite aperture, the eiconal GO approximation is broken and the diffraction effects become essential when a beam propagates in medium, so the geometric ray has to be replaced by the so-called diffraction-ray (DR) [34, 36]. Generally, each DR follows a curvilinear trajectory, which is an integral curve of the transverse spatial component of the Pointing vector. Owing to its definition, DRs do not intersect each other in contrast to GO-rays.

Following [32], we write out the basic formulae of diffraction-ray optics. The equation describing the evolution of DR transversal coordinate $\mathbf{R}_d$ is derived from the parabolic equation for complex amplitude of optical field (see, Eq. (A1)) and has the following form:

$$\frac{d^2 \mathbf{R}_d}{dz^2} = \frac{1}{2\varepsilon_0} \nabla_\perp \varepsilon_{ef} \quad (4)$$

Here, $\varepsilon_{ef}$ is the effective dielectric permittivity of the propagation medium [31], which can be represented as the sum of three components:

$$\varepsilon_{ef} = \varepsilon_0 + \varepsilon_N + \varepsilon_d, \quad (5)$$

with $\varepsilon_0$ representing the nonperturbed medium permittivity (including initial wavefront curvature), and $\varepsilon_N$ being the additive accounting for nonlinear wave refraction caused by the optical Kerr effect and refraction by self-induced plasma. The term $\varepsilon_d = \Delta_\perp A / k_0^2 A$ stands for the diffraction spreading, where $A$ is the real amplitude of optical field, and $k_0 = 2\pi/\lambda_0$ is its wavenumber in vacuum. In Eqs. (4) and (5) we follow the standard representation of a complex field $U$ through the real-valued and slowly-varied amplitude $A$ and phase $\varphi$: $U = A \exp\{i\varphi\}$.

In stationary conditions, one can link the effective permittivity of medium $\varepsilon_{ef}$ with the optical wave phase $\varphi$, and instead of Eq. (4) use a more suitable equation for calculating the diffraction-ray trajectory:

$$\frac{d\mathbf{R}_d}{dz} = \frac{1}{k_0} \nabla_\perp \varphi. \qquad (6)$$

This equation defines a direct link between a diffraction ray and the direction of optical energy flux in medium given by the transverse Pointing vector $\mathbf{S}_\perp$: $\nabla_\perp \varphi = \mathbf{S}_\perp k_0 / A^2$. And hence:

$$\frac{d\mathbf{R}_d}{dz} = \frac{\mathbf{S}_\perp}{A^2}. \qquad (7)$$

Note that because of time averaging of the parabolic wave equation Eq. (A1) over the optical pulse length, we will further consider the time-averaged DRs.

Consider the DR trajectories obtained through the numerical solution to Eq. (A1) and Eq. (6) for synthesized corona-beams. DRs are shown in Figs. 5(a, b) by differently colored groups of curves for two corona-beam profiles (CB$_4$ and CB$_{12}$) with equal initial power $\eta = 10$. Each DR group defines a spatially localized light structure - a diffraction-ray tube [34, 37]. In contrast to an infinitely thin optical beam, a diffraction-ray tube is characterized by a finite cross-section $\sigma_t(z)$ and carries a certain (non-zero) amount of light power (energy) $Q_t(z)$. The fundamental property of a DRT is the invariance of light power (energy) flowing through any of its cross-section, if there are no field sources and sinks in the tube.

Worthwhile noting, in the physical analysis of radiation propagation the DRT model provides a logical delimiting of individual energy flows within the optical beam. As these energy flows are isolated from each other and do not exchange energy (DRTs do not intersect), each of them actually represents a separate sub-beam with the parameters of effective transverse size $R_t = \sqrt{\sigma_t/\pi}$ and coordinate of beamlet centroid $\mathbf{r}_t(z)$ accounting for DRT angular divergence.

Similar to usual light beams, the conservation law for total energy is implemented within every DRT, and the average (effective) parameters can be used to describe the tube evolution. Thus, the DRT cross-section area is subject to the following generalized parabolic law:

$$\frac{d^2\sigma_t}{dz^2} = \frac{1}{\pi Q_t \varepsilon_0} \int_{\sigma_t} \{(\nabla_\perp \varepsilon_{ef} \cdot \mathbf{r}_\perp) I\} d^2\mathbf{r}_\perp, \qquad (8)$$

Formally, this equation resembles the governing DR equation Eq. (4) providing that the averaging of the gradient of medium effective permittivity $\nabla_\perp \varepsilon_{ef}$ is calculated over the DRT cross-section weighted with the local tube intensity $I(\mathbf{r}_\perp, z)$.

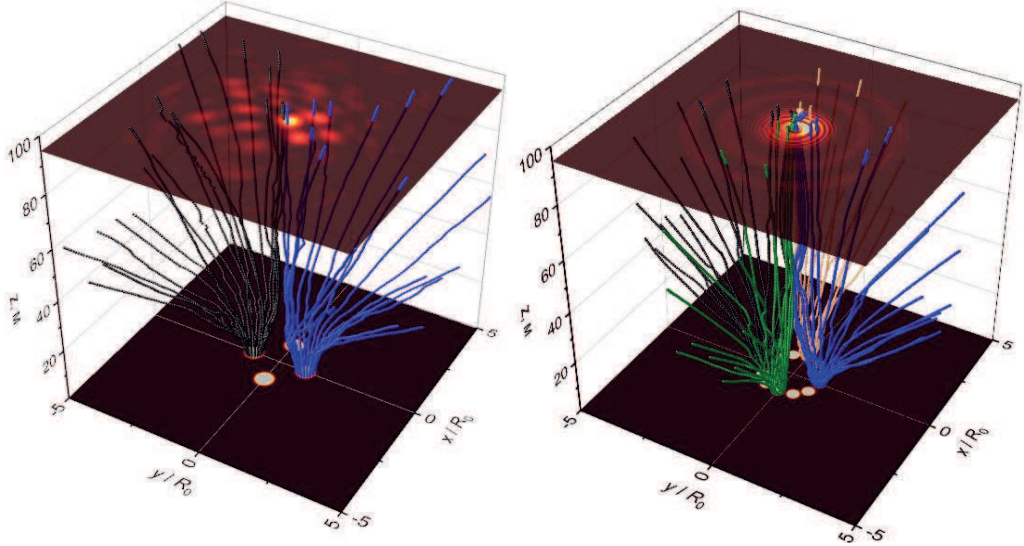

Fig. 5. 3D-evolution of the diffraction-ray tubes during the filamentation of synthesized corona-beams (a) $CB_4$ and (b) $CB_{12}$ with $\eta = 10$. Note, only certain DRTs are shown for better readability.

Consider Fig. 5(a) showing the dynamics of two DRTs formed during the filamentation of synthesized beam with four beamlets ($CB_4$). The diffraction rays which build the outer boundary of each tube are highlighted in the pictures in different colors for better visibility. 2D-distributions of beam normalized optical intensity are also presented in the picture for two distances $z = 0$ m and 95 m.

It is clear that at the initial stage of beam propagation, approximately up to the nonlinear focus located at $z = 7$ m, both plotted DLTs evolve independently of each other. Meanwhile, they slightly change in size, maintaining a circular cross-section shape that indicates light field self-channeling within tubes due to the diffraction suppression by Kerr self-focusing. In this case, the filamentation and plasma generation appear only within the inner regions of each tube (see Fig. 3(b)).

After the nonlinear focus, when the filamentation stops, the DRT cross-sections sharply increase and their spatial shape is distorted. The tubes are stretched in the direction towards beam periphery, and near the optical axis one can realize a crowding of the diffraction rays constituting the tube boundary. This ray crowding indicates a diffraction coupling of neighboring DRTs through the optical phase because no energy exchange between DRTs is possible. Eventually, isolated diffraction peaks in the far-field region are formed.

In the synthesized beam of coronary profile with a large number of sub-apertures ($CB_{12}$ in Fig. 5(b)), the evolution of the diffraction-ray tubes differs from the above-considered case first of all by the fact that there is no noticeable region with the self-similar propagation of individual corona-spikes. Due to power shortage ($\eta_b < 1$) in the sub-beams, the conditions for filamentation in the tubes are not fulfilled and DRTs demonstrate a marked diffraction spreading practically from the beginning of beam propagation. At the same time, a part of diffraction rays at the borders of individual tubes are gathered at the optical axis where they form a new DR family. This DR family represents an axial maximum of intensity and propagates quite independently demonstrating sinusoidal ray trajectories which are typical for a specific class of GRIN (gradient refractive index) optical elements known as a SELFOC fiber [38].

By analogy with a SELFOC, where a negative transverse gradient of *material* dielectric permittivity $\nabla_\perp \varepsilon < 0$ in the waveguide is fabricated, in the case considered here the confinement of DRs near the beam center is maintained due to wave diffraction in negative gradient of the *effective* dielectric permittivity, $\nabla_\perp \varepsilon_{ef} < 0$, which is self-induced under the action of a specific beam profile and nonlinearity of the propagation medium [14]. Thus, at certain propagation ranges when DR crowding within central rays family occurs, the field intensity periodically increases that starts the recurrent beam filamentation.

The dynamics of DRT cross-section area along the propagation distance in normalized values $\sigma_t/\sigma_{t0}$, where $\sigma_{t0}$ is the initial area of the ray tube cross-section, is shown in Figs. 6(a, b). The calculations are carried out based on Eq. (8) provided by the preliminary calculated amplitude and phase profiles of optical field as a result of the numerical solution of parabolic wave equation Eq. (A1). As a reference, the ratio of ray tube area $\sigma_t/\sigma_{t0}$ is also shown here for the case of linear propagation (marked by "LP" in the figures) of the synthesized beams with reduced power $\eta \ll 1$.

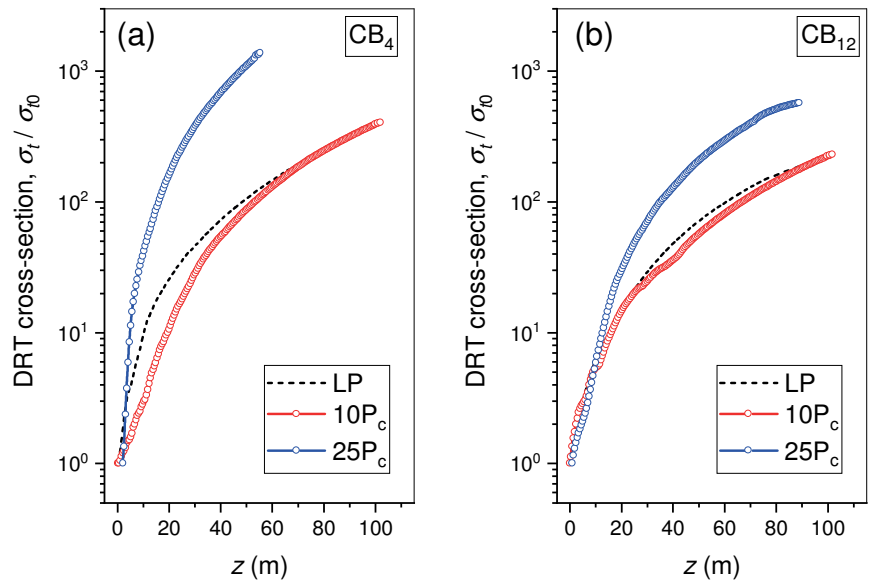

Fig. 6. Dynamics of relative DRT cross-section area upon synthesized beams propagation in filamentation mode in air for (a) $CB_4$ and (b) $CB_{12}$. *LP* denotes the corresponding dependence for linear diffraction regime.

After inspection of these figures, several points can be revealed. First, according to the parabolic nature of Eq. (8), the cross-sectional area of DRT always increases with propagation distance even in the absence of medium optical nonlinearity. In addition, the tube area growth rate increases with the increase of pulse power η analogously as it happens with the beam as a whole [39]. Finally, from the comparison with linear propagation regime, one can see the manifestation of the effects of (a) diffraction compression by the Kerr self-focusing in the sub-apertures in $CB_4$ and (b) the partial power shortage in $CB_{12}$ for beam power η = 10. In the first case, the DRT area is smaller than that in the absence of medium nonlinearity (LP-mode), while in the second case, to the contrary, the tube cross-sections in linear and nonlinear regimes are almost equal up to the nonlinear focus.

## 5. "Dressed" corona-beam filamentation

In the previous sections we show that the amplitude distribution of corona-beams in both linear and nonlinear propagation regimes always demonstrates the formation of an extra maximum in the beam centroid, i.e., on the beam optical axis. This is caused by the diffraction spreading of individual sub-apertures occurring even under the influence of strong Kerr self-focusing. As can be seen in Fig. 5(b), regardless of self-compression of the sub-beam central part in the Kerr-medium, the peripheral beam areas are always subject to the diffraction spreading rather than compression. Because of this angular ray dispersion, the edge DRs begin crowding at the meeting point in the central beam area that indicates field interference of individual sub-beams and the formation of a new diffraction maximum at the synthesized beam center. Starting from this distance, the spatially delayed beam filamentation can initiate if there is a power shortage in the partial sub-apertures, or recurrent filamentation if such a shortage is absent. The question arises: Can one manipulate the central diffraction maximum and consequently the dynamics of the filamentation of the corona-beams?

Our study shows that the most effective way of such control is to create this central lobe *a priori* within the initial coronary profile by transforming a CB into a "dressed" CB (DCB) in the terminology adopted above. For the sake of simplicity, in the simulations we fixed the transverse size of this central extra sub-beam at some value and vary only its amplitude $U_d$ as a free parameter in the intensity distribution Eq. (3). Hereafter, a dimensionless ratio $\delta = (U_d/U_{cb})^2$ will be used to characterize DCBs with different profiles. This parameter shows the fraction of energy (or power) containing in the central spike compared to any single peripheral sub-beam.

The dynamics of maximal pulse intensity at DCB nonlinear propagation with some fixed number of ring sub-apertures ($N_b = 8$) but with different power fraction δ is shown in Fig. 7(a). As seen, the increase in energy content of the central lobe leads to the appearance of filaments at the very beginning of laser beam propagation. It is clear that this filamentation is realized in the DCB center. However, the most interesting fact is that the presence of the central sub-beam can significantly delay the distance of consecutive (secondary) filamentation. Indeed, at δ = 2, the re-filamentation of DCB occurs at the distance of $z_{end}$ = 179 m instead of 125 m in the absence of a central maximum (δ = 0).

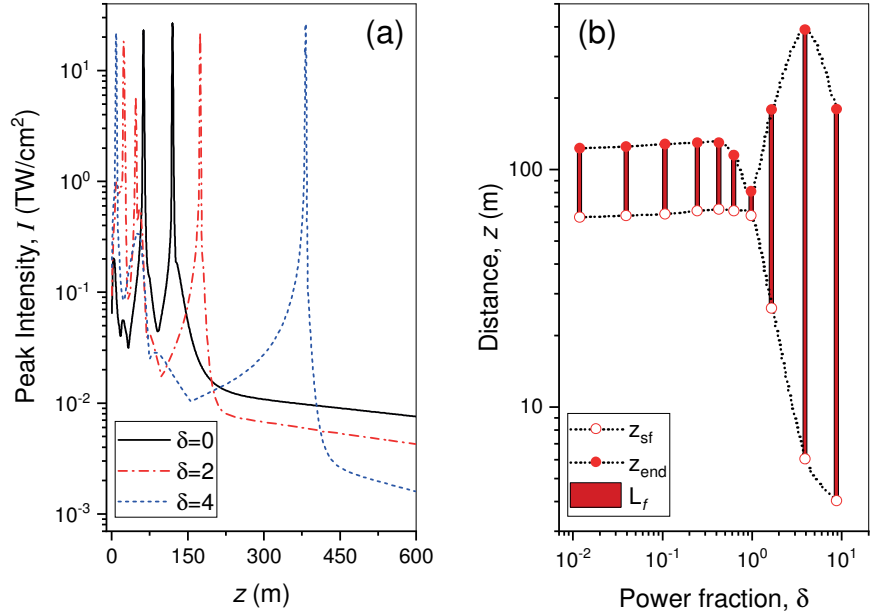

Fig. 7. (a) Beam intensity trace evolution and (b) filamentation parameters for "dressed" corona-beams ($N_b = 8$, $\eta = 10$) depending on the central spike power fraction $\delta$.

It turns out that this filamentation delay can be controlled within certain limits by increasing or decreasing the ratio $\delta$. The result of such control is demonstrated in Fig. 7(b), where the dependencies of the starting $z_{sf}$ and ending $z_{end}$ coordinates of the filamentation region for different DCBs with equal initial reduced power $\eta$ are presented. The values of the total filamentation length $L_f$ are shown by bold bars in this figure also. It follows, that at rather small power fraction of the central sub-beam ($\delta < 0.5$), its influence on the parameters of filamentation region is insignificant. However, at larger values of $\delta$ the effect of the central sub-beam becomes pronounced.

First, at moderate power fraction, $\delta \approx 1$, the range $z_{end}$ of DCB re-filamentation decreases and the total filamentation length shortens from $L_f \approx 60$ m to 17 m. If the central extra sub-aperture becomes leading ($\delta > 1$), the specific filamentation scenario of the "dressed" beam is realized [14], when the role of the annular low-energy area is not only the energy replenishment for medium ionization losses by the filamented central maximum, but also the formation of a specific "diffraction waveguide" which promotes the self-channeling of this central beam part and its concentration within the boundaries of the filament. As seen below, the stability and length of this "diffraction waveguide" are mostly affected by the pulse energy of the circular region of the DCB.

## 6. Dynamics of plasma-free post-filamentation propagation of synthesized corona-beams

After the end of filamentation, the laser-induced plasma generation in the medium becomes weak or practically stops, while the beam transverse profile remains strongly heterogeneous and in some sense can be treated as a spatially-structured light [40]. Inside the laser beam, several localized light structures (bright spots) continue to exist at sufficiently long ranges. Some of these light structures, which will be referred hereafter as the post-filamentation light channels (PFCs), possess enhanced intensity (up to several TW/cm$^2$) and lowered angular divergence in comparison with the laser beam as a whole. Thus, the first experiments of TERAMOBILE group [6] revealed the backscattered radiation from the PFCs at ranges of about 20 km upon vertical propagation of TW-power laser pulses in the Earth atmosphere. In [41] this "new propagation regime without ionization" in the form of post-filamentation channels is experimentally investigated and first theoretical explanation is presented.

Currently, the physical model of a PFC treats it as a specific light structure, which is self-organized inside the beam after the filamentation termination, as a result of optical pulse diffraction on the plasma bunch formed in the beam wake supported by the continuing focusing action of Kerr nonlinearity in the high-intensity areas of the pulse [42-44]. The scientific and practical interest in the post-filamentation pulse self-channeling is conditioned by the prospects of PFC application as a means of high-intensity optical radiation delivering over long ranges in the atmosphere [45]. In this connection, below we consider some specific features of PFC formation by coronary profile optical beams.

Recall Fig. 2 and Fig. 4 showing the transverse distribution of the relative pulse intensity in CBs with different number of sub-apertures along the propagation path. It is clearly seen that for all types of intensity distribution, a single or several leading maxima of intensity are formed starting from certain distance. Later on, at distances exceeding the filamentation range, the beam profile always exhibits a central bright spot, a PFC, surrounded by low-intensity rings. This PFC increases in size and decreases in intensity upon propagation under the influence of field diffraction being partially counter-balanced by Kerr self-focusing. Meanwhile, the rate of this diffraction broadening differs for the CBs with different profiles.

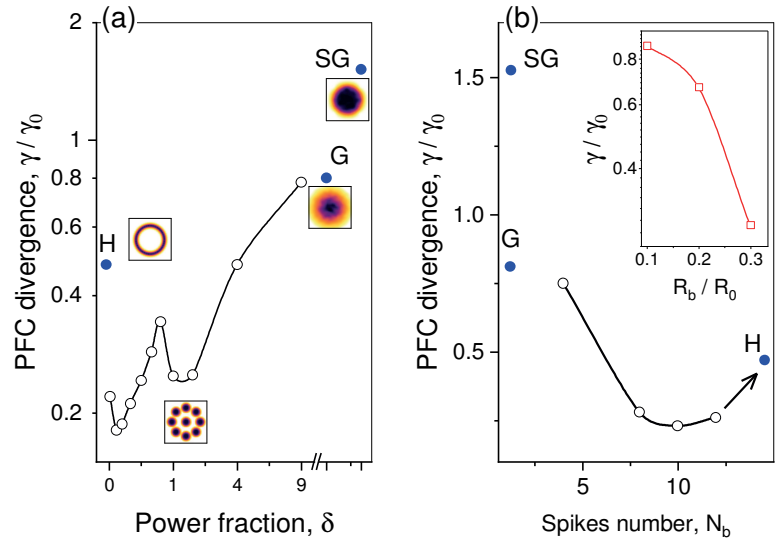

Fig. 8. PFC angular divergence for different "dressed" corona-beams ($N_b = 8$, $\eta = 10$) as a function of (a) power fraction $\delta$ and (b) sub-beams number $N_b$. Inset in (b): PFC divergence depending on sub-aperture radius $R_b$ ($N_b = 8$). «H» denotes the dark-hollow beam, «G» is a Gaussian beam, and «SG» is a super-Gaussian beam profile.

Dynamics of PFC angular divergence γ acquired to the distance $z = 500$ m by DCBs with different central spike power fraction δ, size $R_b$ and annular sub-beams number $N_b$ is shown in Figs. 8(a, b). The calculations are carried out using the evolutionary equation (8) for the central DRT by the following expression: $\gamma = \left(2\sqrt{\pi}\right)^{-1}\sqrt{d^2\sigma_t/dz^2}$, $z > z_{end}$. Here, the PFC divergence γ is normalized to the natural diffraction divergence of a Gaussian intensity distribution: $\gamma_0 = (kR_0)^{-1}$.

From the inspection of Fig. 8(a) it follows that the PFC divergence γ can be manipulated within considerable limits by changing the energy fraction contained in the central sub-aperture of DCB. In this case, in comparison with the beams of unimodal profiles, such as Gaussian (G), super-Gaussian (SG) and dark-hollow (H) transverse intensity distributions, the PFC in the corona-beams always maintains a higher energy concentration along the propagation distance. The values of angular divergence γ for unimodal-profile beams are also shown in this figure by blue solid circles.

In the range of relatively low central spot energy $0.01 < \delta < 1$, the angular divergence of the PFC exponentially increases, while according to Fig. 7(b) there are no essential changes in position and length of filamentation area. At $\delta \approx 1$, the PFC divergence demonstrates first an extremum, then it sharply decreases practically to the value realized for DCB without central sub-aperture ($\delta = 0$), and finally γ begins increasing monotonically. This stable growth of γ is directly related to the weakening of the confinement ability of the SELFOC- waveguide appearing in DBC annular area as its energy content decreases (δ is increased). As a result, the central PFC is less affected by the surrounding circular field structure and experiences a stronger diffraction divergence. Interestingly, the growth in γ at the increasing δ correlates here with the decreasing of the starting coordinate $z_{sf}$ of beam filamentation (Fig. 7(b)). This once again suggests the fact that for achieving a PFC with minimal angular divergence at a given distance, it is necessary to initiate the beam filamentation as far as possible from the optical path beginning.

The number of partial sub-apertures in DCB as well as their size has strong influence on the angular divergence of the emerging post-filamentation central light channel. Indeed, as Fig. 8(b) shows, in the four-modal CB ($N_b = 4$) the formed PFC has a divergence close to that of a Gaussian beam, $\gamma/\gamma_0 \approx 0.75$, with the same initial power η. Increasing the number of sub-apertures up to nine ($N_b = 8$ plus the central sub-beam) sharply reduces, practically by three times, the PFC angular divergence ($\gamma/\gamma_0 \approx 0.27$). The corona-beam with ten annular sub-beams and one central spike forms PFC of minimal divergence: $\gamma/\gamma_0 = 0.23$. However, further increase of the sub-beams number $N_b$ again causes PFC divergence γ growth, which in the limit $N_b \to \infty$ tends to the value of a hollow beam divergence: $\gamma/\gamma_0 = 0.48$.

The Inset to Fig. 8(b) shows the PFC divergence for DCB with $N_b = 8$ depending on the initial radius of partial sub-beams $R_b$. As expected, γ monotonically decreases if the sub-beam size increases because the self-focusing distance of both beamlets and the beam as a whole increases as the sub-beams radius becomes larger (see Fig. 3(c)) providing that the fractional power $\eta_b$ in each sub-beam is fixed. This reduces the diffraction spreading of the formed PFC. Obviously, due to the inherent geometric limitation on the parameter $N_b R_b \leq \lfloor \pi R_0 \rfloor$ of the considered synthesized corona-beam (here $\lfloor \ldots \rfloor$ denotes an integer part), the enlargement of the sub-apertures leads to the reduction of their number in the beam. Therefore, for each value of beam power $\eta$, there should be the optimal set of parameters $N_b$ and $R_b$ providing the minimal PFC divergence. Our simulations show that for $\eta = 5$, 10 and 25 these parameter pairs ($N_b : R_b/R_0$) are as follow: (4:0.4), (10:0.25) and (16:0.15), respectively.

## 7. Conclusions

Based on the results of our numerical study the following conclusions can be made regarding the multiple filamentation dynamics in air of a particular type of high-power synthesized laser beams with a coronary intensity distribution, which is constructed by incoherent combining of individual sub-beams of smaller aperture arranged in a circular ring.
- (a) In contrast to the beams with a unimodal intensity distribution, such as Gaussian and super-Gaussian (plateau-like), the filamentation of corona-beams are characterized by striking manifestation of beam diffraction and Kerr self-focusing counteraction over the spatial scales of individual sub-apertures (Figures 2 and 4). As a result, the coordinates of nonlinear foci, where the recurrent filamentation occurs, can be significantly shifted away from the start of corona-beam propagation (Fig. 3(a));
- (b) The onset distance and spatial extent of CB filamentation region can be effectively controlled within a wide range by changing the number and arrangement of corona spikes in the synthesized coronal profile (Figs. 3(c) and 7(b));
- At the post-filamentation stage of corona-beam propagation, the angular divergence of its most intense central region (PFC) is usually lower than that of a unimodal (e.g., Gaussian) beam profile. The value of PFC divergence in a corona-beam significantly depends on the specific CB profile (Fig. 8). Our simulations show that for each initial value of CB power, the optimal correlation of the number, radius and fractional power of corona sub-beams can be found providing the minimal angular divergence of the produced PFC.

## 8. Funding

The work was partially supported by the Ministry of Science and Higher Education of the Russian Federation and by the Russian Scientific Foundation (Agreement #21-12-00109).

# Appendix

For the simulation of multiple filamentation of cm-wide aperture laser corona-beams, we utilize the reduced 3D-version of NLSE, which is obtained by integration of the full (3D+1) NLSE with respect to the temporal dimension. Being inherently approximate, this integrated-in-time model is still capable to describe the principal features in the experimental patterns and cardinally simplifies the calculations.

To obtain the equation for the mean (in time) filed amplitude $U(\mathbf{r}_\perp,z) = (1/2T)\int_{2T} \tilde{U}(\mathbf{r}_\perp,z,t)dt$ ($T$ is the temporal grid size), the temporal profile of the envelope $\tilde{U}$ is fixed in the stepwise form with some constant in $z$ pulse half-width $t_0$. This means that the pulse temporal shape is considered to be "frozen" in space, so one can set:

$$\tilde{U}(\mathbf{r}_\perp,z,t) = \begin{cases} U_0 U(\mathbf{r}_\perp,z), -t_0 \leq t \leq t_0 \\ 0 \end{cases}, \qquad (A1)$$

Here, $U(\mathbf{r}_\perp,z)$ is a function of only transverse coordinates $\mathbf{r}_\perp$ and evolutional variable $z$, and $U_0 = \sqrt{\pi I_0} t_p / 2t_0$ is the amplitude. Usually, the value of $t_0$ should be properly chosen in the range $0 < t_0 \leq t_p$ ($t_p$ is initial pulse duration) to provide the best agreement with the results of the full (3D+1)-model calculations. In our case, the value $t_0 = 0.1 \cdot t_p$ is proved to accounting for pulse temporal compression due to Kerr self-focusing.

Now one can consider the integral NLSE neglecting the pulse group velocity dispersion effect, the inertia of the Kerr nonlinearity, and the transient nature of medium photoionization dynamics. Within these approximations, we obtain the following 3D-equation for the time-averaged electric field amplitude $U(\mathbf{r}_\perp,z)$:

$$\left\{ \frac{\partial}{\partial z} - \frac{i}{4}\nabla_\perp^2 - i\frac{L_R}{L_K}|U|^2 + \frac{L_R}{2L_W} W_I \frac{(1-B)}{|U|^2} + \frac{iL_R}{2L_{pl}}\left(1 - \frac{i}{\omega_0 \tau_c}\right) B \right\} U = 0, \qquad (A2)$$

Here, $\nabla_\perp^2$ is the transverse Laplacian, $\omega_0 = 2\pi c/\lambda_0$ is the central pulse angular frequency, $\tau_c$ is the characteristic electron collision time. This equation introduces the characteristic lengths of Kerr $L_K$ and plasma $L_{pl}$ nonlinearities, as well as the multiphoton ionization length $L_W$ in the propagation medium [29]. The coefficient $B$ takes into account the total (over the pulse) gain of free electrons in the beam wake upon multiphoton and impact ionization of air molecules. The multiphoton photoionization rate $W_I$ is calculated using the Perelomov-Popov-Terent'ev model [46], which is applied to a gas mixture simulating atmospheric air (20%$O_2$+80%$N_2$) at the total neutral molecule concentration of $2.5 \cdot 10^{19}$ cm$^{-3}$. The critical self-focusing power $P_c$ of laser radiation at the wavelength of 800 nm in air is set as 3.2 GW [47].

In numerical calculations, the initial transverse distribution of the laser radiation amplitude possesses a plane wavefront and is set in Descartes coordinates $\mathbf{r}_\perp \equiv (x,y)$ according to the expressions (1)-(3). Additionally, the realistic beam quality degradation inherent to powerful laser sources is simulated by superimposing a random amplitude noise on the generated smooth intensity profile as follows: $U_\perp(x,y) \to U_\perp(x,y)\left[1 + A_m \cdot \tilde{f}(x,y)\right]$, where $\tilde{f}$ denotes a random variable normally distributed in the range $[-1...1]$, and $A_m$ is noise amplitude. To be specific, all calculations are performed for laser beams with the initial $1/e$ radius $R_0 = 5$ mm and pulse duration $t_p = 100$ fs. The numerical grid dimensions in spatial coordinates are 20 cm × 20 cm with $2^{13} \times 2^{13}$ grid points. The numerical scheme is advanced along $z$ coordinate using an adaptive step.

To confirm the validity of the reduced 3D-NLSE in the simulation of laser beam filamentation with a structured spatial profile, we carried out test calculations of the nonlinear propagation in air of corona-beam with different power using the full-scale UPPE equation for complex envelope of the electric field $U_{k\omega} \equiv U(k_\perp, z; \omega)$ in the spatiotemporal frequency domain [48]:

$$\frac{\partial U_{k\omega}}{\partial z} = i\left(k_z - \omega_0/\upsilon_g\right)U_{k\omega} + i\frac{\omega^2}{2c^2 k_z}\frac{P_{k\omega}}{\varepsilon_0}, \tag{A3}$$

Here $k_z = \sqrt{k(\omega)^2 - k_\perp^2}$ is the propagation constant along $z$, $k_\perp^2$ is the squared transverse wavevector component, $k(\omega) = \omega n(\omega)/c$ is the wavenumber depending on the pulse angular frequency $\omega$ and the chromatic dispersion of air refraction index $n(\omega)$, $\varepsilon_0, c$ stay for dielectric permittivity and light speed in vacuum, $P_{k\omega}$ is the medium nonlinear polarization which accounts for all significant physical mechanisms causing pulse self-phase modulation and nonlinear energy dissipation during the propagation [49]. Eq. (A3) is formulated in the pulse coordinate frame with the origin moving with pulse group velocity $\upsilon_g = (\partial k/\partial \omega)^{-1}$.

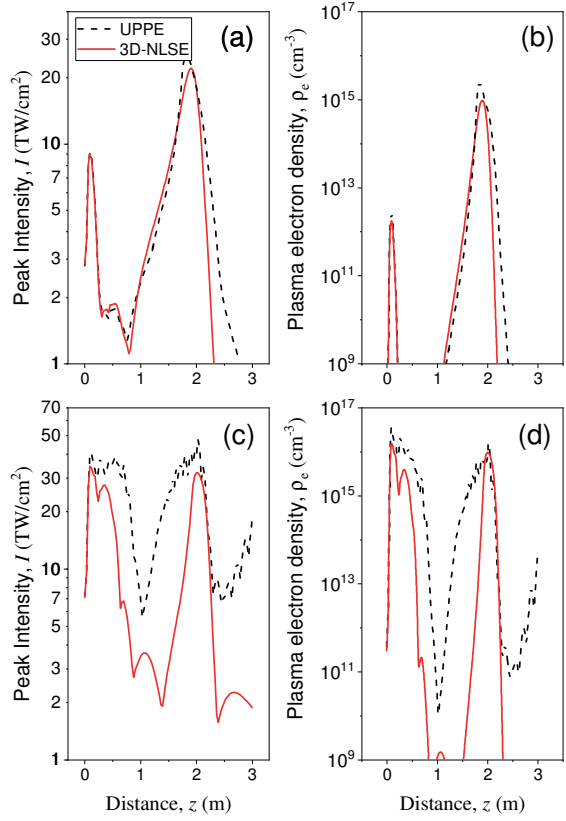

Fig. A1. (a, c) Pulse peak intensity $I$ and (b, d) free-electron volume density of plasma $\rho_e$ during corona-beam multiple filamentation with $\eta = 10$ (a, b) and 25 (c, d). The calculations are carried out using 3D-NLSE Eq. (A1) and UPPE Eq. (A4).

Figure A1 demonstrates the comparison of the filamentation dynamics of a synthesized corona-beam carried out by means of the 3D-NLSE and 4D-UPPE. The calculations are performed for a coronal beam profile given by Eq. (1) combined by ten partial sub-beams ($N_b = 10$). Here, the initial CB radius is set to $R_0 = 1$ mm to reduce the processor time of the 4D-UPPE integration to a realistic value. Accordingly, the beam diffraction length $L_R$ (Rayleigh range) is also reduced to the value of 62 cm. In the conditions when the calculation domain is discretized by approximately $2^{40}$ grid nodes and the task is run on 48 processor cores of a supercomputer cluster [50] based on blade servers with Intel Xeon E5-2680v3 processors, the total run time is about 30 hours.

From the analysis of Figs. A1(a-d) it follows that the reduced 3D-model Eq. (A1) with satisfactory accuracy predicts both the appearance of diffraction maxima in the synthesized beam with low power at $z \approx 18$ cm (Fig. A1(a)) and the formation at longer distances the nonlinear focuses with the plasma density characteristic for laser filaments (Figs. A1(b, d)). The description validity of the CB filamentation dynamics slightly degrades with the increasing pulse power, but the length and location of the filaments are predicted quite correctly by the reduced 3D-NLSE version.